# Bragg-like microcavity formed by collision of single-cycle self-induced transparency light pulses in a resonant medium

ROSTISLAV ARKHIPOV[1,2*], ANTON PAKHOMOV[1], OLGA DIACHKOVA[1,2], MIKHAIL ARKHIPOV[1,2], NIKOLAY ROSANOV[1,2]

[1]*St. Petersburg State University, St. Petersburg, 199034 Russia*
[2]*Ioffe Institute, 194021 St. Petersburg, Russia*
*\*arkhipovrostislav@gmail.com*

**The coherent interaction of extremely short light pulses with a resonant medium can result in formation of population difference gratings. Such gratings have been created by pulses that are π/2 or smaller. This paper demonstrates that a microcavity with Bragg-like mirrors can be formed by colliding two single-cycle attosecond self-induced transparency pulses in the center of a two-level medium. The parameters of this structure can be quickly adjusted by increasing the number of collisions, which showcases the ability to control the dynamic properties of the medium on a sub-cycle time scale by using attosecond pulses.**

## 1. Introduction

Electromagnetically induced gratings are the result of interference between two monochromatic laser beams in a medium [1]. Among the most important types of such gratings are fiber Bragg gratings which are widely used for the dispersion compensation or as filters in fiber optics [2]. In different configurations, the interference light pattern creates a periodic distribution of the atomic level population in the medium. Gratings made using this method have a wide range of applications, including spectroscopy [1], the creation of all-optical modulators [3], the measurement of ultra-short UV pulses [4], and others, as described in [5]. However, such an approach does not allow for fast control (creation, erasure, and spatial period change), which is critical. The idea of controlling such gratings with extremely short pulses is reasonable; the shorter the pulse, the faster the medium properties can be changed.

Advances in the generation of extremely short femtosecond and attosecond electromagnetic pulses [6] have enabled the study and ultrafast control of wavepacket dynamics in atoms, molecules, and solids [7-8], paving the way for the development of ultrafast optoelectronics [9]. The production of single-cycle pulses, which contain two half-waves of opposite polarity [7], and unipolar half-cycle pulses, which contain a single half-wave, represent the limit for reducing the duration of electromagnetic pulses in a given spectral range [10]. So far, single and half-cycle extremely short attosecond pulses (ESPs) have been observed experimentally [7,11] and theoretically proposed [10,12,13]. The review [10] summarizes recent findings on the generation of ESPs and their interaction with matter.

Another example of ultrafast control of the medium's state using ultra-short pulses is the generation and control of population difference gratings, microcavities, and polarization waves in a resonant medium [4, 14-16]. In particular, it can be applied in ultra-high resolution holography of fast moving objects [17]. Such structures can be created with or without pulse overlap in the medium. A review of recent results in this area can be found in [4,18] and the referenced literature.

These gratings result from an ESP's coherent interaction with the medium when the pulse duration and delay between them are less than the medium's polarization relaxation time $T_2$. The explanation for the grating's formation is the interference of polarization wave induced by the previous pulse with the each subsequent pulse entering the medium [4,14-16].

In our first studies [14-16,18], the gratings were induced by ESPs acting as π/2, π (or even less) pulses, which transferred the medium from the ground state to a state with a zero population difference. These gratings had a harmonic shape and filled the entire medium. In [19], the collision dynamics of unipolar 2π-solitons of self-induced transparency (SIT) was considered in a medium consisting of a mixture of amplifying and absorbing particles.

The formation of a light-induced microcavity by the self-action of a single-cycle pulse, leading to its self-stopping, has been demonstrated in [20]. The possibility of the inducing and shaping dynamic microcavties during the collision of rectangular unipolar pulses in a two-level resonant medium has been demonstrated [21-23]. In some cases, a population difference jump occurred in the center of the medium - in the pulse collision region, while the medium remained non- excited at the sides.

In this work, the dynamics of the population difference gratings and polarization upon the collision of single-cycle light pulses, which act in a similar way to self-induced transparency (SIT) pulses [24], is studied. The amplitude and duration of the pulses were chosen so they act like 4π-pulses: i.e. each half-wave of a single-cycle pulse behaves like a separate 2π SIT pulse. It is shown that an unusual phenomena arises in the medium, namely the formation of a microcavity surrounded by the Bragg-like mirrors. In the medium's center there is a narrow region where the population difference is 1 (the medium is not excited). At the sides, gratings of the population difference (Bragg structures) are formed. Otherwise, the medium remains non-excited.

It should be noted that these studies are also relevant to the ongoing research of the properties of media with a rapidly changing refractive index, both in space and in time - space-time photonic crystal (SPC), as reviewed in [25,26]. In our case, a sequence of ESPs causes a rapid change of state in the medium, resulting in the formation of population gratings.

## 2. Results of numerical simulations

To study the grating dynamics for the medium inversion and polarization, we numerically solve the system of Maxwell-Bloch equations for the non-diagonal density matrix element of the two-level medium $\rho_{12}$, the population difference (inversion) $n = \rho_{11} - \rho_{22}$ of the two-level medium, its polarization P together with the one-dimensional wave equation. This system of equations is written as [27]

$$\frac{\partial \rho_{12}(z,t)}{\partial t} = -\frac{\rho_{12}(z,t)}{T_2} + i\omega_0 \rho_{12}(z,t) - \frac{i}{\hbar} d_{12} E(z,t) n(z,t) \quad (1)$$

$$\frac{\partial n(z,t)}{\partial t} = -\frac{n(z,t) - n_0(z)}{T_1} + \frac{4}{\hbar} d_{12} E(z,t) \mathrm{Im} \rho_{12}(z,t), \quad (2)$$

$$P(z,t) = 2N_0 d_{12} \mathrm{Re} \rho_{12}(z,t), \quad (3)$$

$$\frac{\partial^2 E(z,t)}{\partial z^2} - \frac{1}{c^2} \frac{\partial^2 E(z,t)}{\partial t^2} = \frac{4\pi}{c^2} \frac{\partial^2 P(z,t)}{\partial t^2} \quad (4).$$

This system of equations contains the following parameters: $t$ – time, $z$ – longitudinal coordinate, $c$ – speed of light in vacuum, $N_0$ – density of two-level atoms, $\hbar$ is the reduced Planck constant, $n_0$ – population difference in the absence of the electric field ($n_0 = 1$ for an absorbing medium). Values of other parameters are given in the table below.

The system of equations (1)-(4) was solved numerically using the parameters given in the Table 1. The wave equation was solved by the time-domain finite-difference method, the equations for the density matrix were solved by the 4th order Runge-Kutta method.

Initially, a pair of single-cycle pulses 1 and 2 were injected into the medium from left to right and right to left, respectively, as:

$$E(z = 0, t) = E_{01} e^{-\frac{(t-\tau_1)^2}{\tau^2}} \sin[\omega_0(t - \tau_1)], \quad (5)$$

$$E(z = L, t) = E_{02} e^{-\frac{(t-\tau_2)^2}{\tau^2}} \sin[\omega_0(t - \tau_2)], \quad (6)$$

**Table 1. Problem parameters used in the numerical calculation**

| Description | Value |
| --- | --- |
| Wavelength of the resonant transition | $\lambda_0 = 700$ nm |
| Transition dipole moment | $d_{12} = 20$ Debay |
| Population difference relaxation time | $T_1 = 1$ ps |
| Polarization relaxation time | $T_2 = 0.5$ ps |
| Atomic density | $N_0 = 5 \cdot 10^{14}$ cm$^{-3}$ |
| Driving field amplitude | $E_0 = 333\,500$ ESU |
| Driving pulse duration $\tau$ | $\tau = 777.8$ as |
| Delays | $\tau_1 = \tau_2 = 2.5\tau$ |

A sequence of pulses has been generated with the use of zero boundary conditions. Fig. 1 shows the direction of propagation and the time of entry into the medium. This allowed the pulses to come out of the medium, to bounce off the boundaries of the integration region and to propagate back into the medium. The pulses collided at the center of the medium each time. This meant that a sequence of exciting pulses could be generated. The medium was located between the points $z_1 = 4\lambda_0$ and $z_2 = 8\lambda_0$. The full computational domain had a length of $L = 12\lambda_0$. The scheme used to generate the sequence of such pulses is shown in Figure 1b, where all pulses are numbered and their directions of propagation are indicated by arrows.

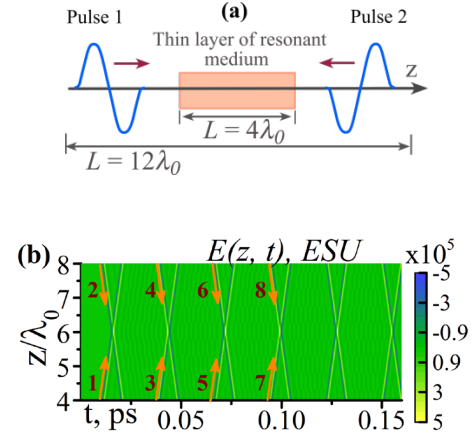

Fig. 1. The diagram showing the direction of propagation of pulses in the medium and their moment of entry.

Figure 2 shows the spatio-temporal dynamics of the population difference (b) and of the induced medium polarization (c) driven by the sequence of pulses propagated according to this scheme. The dynamics of the population difference is unusual. In the region of overlapping pulses in the center of the medium at $z_c = 6\lambda_0$, the medium remains practically unexcited with $n = 1$. At the same time, periodic structures of the population difference are formed to the left and right of the center of the medium, at a small distance from the point $z_c$, i.e. the Bragg reflectors, see Figure 2a. After each collision of the pulses, the spatial frequency of these structures is increased. And in other regions of the medium, away from the region of overlapping pulses, the medium remains in a practically non-excited state: the population difference is close to 1. This is expected because here the propagate in the SIT regime, returning the medium to the ground state.

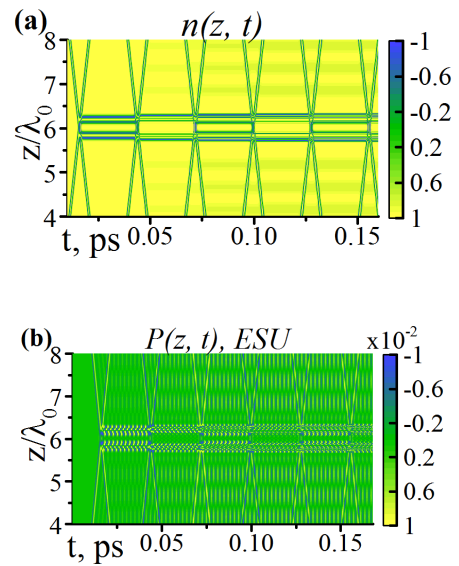

Fig. 2. (a) Dynamics of the population difference n(z,t), (b) dynamics of the polarization P(z,t) under the influence of one-cycle attosecond pulses 1 and 2 of the form (5) and (6), which collide in the center of the medium at the point $z_c = 6\lambda_0$. The table 1 shows the calculation parameters. The medium is located between the points $z_1 = 4\lambda_0$ and $z_2 = 8\lambda_0$. The full integration region had a length of $L = 12\lambda_0$ (the figure shows only the part of this region filled with the medium).

The physical mechanism behind the formation of such microcavities can be explained in the following way. Each half-wave of the pulse behaves as a separate SIT-like soliton. In the center of the medium, at the point $z_c$, where the pulses collide, the medium remains unexcited after they pass through, as well as in the sides of the medium where the pulses do not overlap. However, in the areas to the left and to the right of $z_c$, where the pulses still overlap, there are polarization waves generated by each pulse before the collision, see Fig. 2b. In these regions, the pulses do not act as SIT pulses. They interfere with these polarization waves and coherently control these oscillations. This leads to the appearance of Bragg-like inversion gratings in the pulse overlap region.

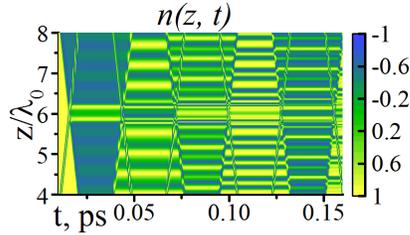

Fig. 3. Dynamics of the population difference n(z,t), $E_{02} = E_{01}/2$. Other parameters are the same as in the Fig.3.

The deviation of the pulse amplitude from the one allowing it to act as a 4π-pulse causes the break-up of the Bragg type structure shown in Fig. 2. To illustrate this, we performed numerical simulations. The amplitude of the second pulse was halved, resulting in a π-pulse, see Fig.3. The microcavity formation in the center is visible as a result of complex striations formed outside the pulse-overlapping region.

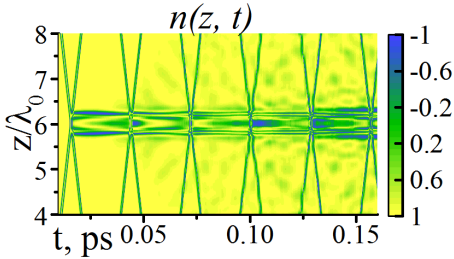

Fig. 4. Dynamics of the population difference n(z,t), $N_0 = 5 \cdot 10^{19}$ cm$^{-3}$. Other parameters are the same as in the Fig.3.

The particle density $N_0$ also affects the structure dynamics. The 4π-pulse changes its shape in the medium when the particle density is high [28]. Fig. 4 shows the dynamics of the population inversion when the particle density was $N_0 = 5 \cdot 10^{19}$ cm$^{-3}$. It can be seen that until the number of collisions is less than 3, the Bragg-type structure is preserved. After a larger number of collisions, the medium does not remain in the ground state in the point $z_c$, and the Bragg-type structure is almost destroyed.

## 3. Bragg-like grating reflectivity

In order to estimate the Bragg-like grating's reflectivity, we will follow the approach from [22]. The population inversion $n(z)$ is related to the index of refraction $n_r$ for a two-level medium as [29]

$$n_r(\omega) \approx \sqrt{1+4\pi\chi_{nr}}\left(1+\frac{1}{\hbar(1+4\pi\chi_{nr})}\frac{4\pi\omega_0 d_{12}^2 N_0(\omega_0^2-\omega^2)}{(\omega_0^2-\omega^2)^2+\frac{4\omega^2}{T_2^2}}n(z)\right) \quad (7)$$

discarding the non-resonant part of the medium susceptibility, i.e. $\chi_{nr} = 0$:

$$n_r(\omega)-1 \approx \frac{1}{\hbar}\frac{4\pi\omega_0 d_{12}^2 N_0(\omega_0^2-\omega^2)}{(\omega_0^2-\omega^2)^2+\frac{4\omega^2}{T_2^2}}n(z) = \Delta n_r(\omega)\cdot n(z) \quad (8)$$

For simplicity, we will consider a uniform Bragg-like 1D grating formed within the medium with an average refractive index $n_0$, where the index of refractive profile can be expressed as

$$n_r = n_0 + \Delta n_r(w)\cdot cos(2\pi z/\Lambda), \quad (9)$$

where $\Delta n_r$ is the amplitude of the induced refractive-index perturbation.

Population difference grating of this type can be formed in areas where pulses do not overlap, as shown in Fig. 3 or, for example, in [21]. The grating's period $\Lambda$ begins with $\lambda_0/2$ and could be multiplied with each consecutive pulse injected, i.e., $\Lambda_2 = \lambda_0/4$, $\Lambda_3 = \lambda_0/8$. ($\lambda_0$ is the resonant transition wavelength). Total reflectivity of harmonic grating can be written as [30]:

$$R(l,\lambda) = \frac{\Omega^2(\sinh sL)^2}{\Delta k^2(\sinh sL)^2+s^2(\cosh sL)^2}. \quad (10)$$

Here $L$ is the grating length, $\Omega$ is the coupling coefficient which can be estimated as $\Omega \cong \pi\Delta n_r/\lambda$, where $\lambda$ is the incident wavelength ($\lambda = \tilde{k}\lambda_0$), $\Delta k = k - \pi/\Lambda$ is the detuning, $k = 2\pi n_0/\lambda$ is the propagation constant, $s = \sqrt{\Omega^2-\Delta k^2}$.

Fig. 5 shows the dependence of the Bragg reflection spectrum at $\Lambda=\lambda_0/2$ as a function of the wavelength $\lambda$ at the different parameters. By changing the length of the grating $L$ and medium particle concentration $N_0$, we observe a non-linear response: with an increase in $N_0$, the width of a grating's reflectivity peaks gets wider, and with a change in $L$, the position of the side lobes gets closer to $\lambda_0$. (Fig.5: from (3) $\lambda_1 = 678.8$ at $L = 4\lambda_0$ to (4) $\lambda_1 = 691.3$ at $L = 10\lambda_0$). In our case, the length of the grating $L$ can be controlled by the delay between pulses or their duration; and by changing $L$ and subsequently the position of secondary reflection peaks we could use such a formation as an easily controllable optical filter.

Comparing Fig. 5 with a typical Bragg reflection spectrum [29], it is clear that the peaks are further apart and their widths are much narrower; this is due to the fact that since our grating is formed on the basis of population inversion of the resonant transition of the medium $\lambda_0$, a pronounced nonlinear dependence on the wavelength $\lambda$ (see eq. (8.)) appears.

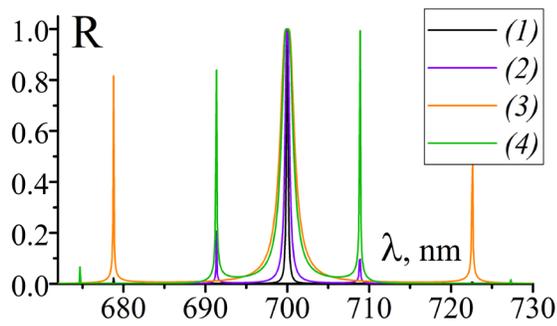

Fig. 5. Reflection $R$ as a function of the wavelength $\lambda$; (1) $L = 4\lambda_0$, $N_0 = 5 \cdot 10^{14} cm^{-3}$ (2) $L = 10\lambda_0$, $N_0 = 5 \cdot 10^{14} cm^{-3}$, (3) $L = 4\lambda_0$, $N_0 = 5 \cdot 10^{19} cm^{-3}$, (4) $L = 10\lambda_0$, $N_0 = 5 \cdot 10^{19} cm^{-3}$. Other parameters are the same as in table 1.

## 4. Conclusions

Thus, in this work an unusual dynamic of the population difference grating and the polarization of the medium upon the collision of a sequence of single-cycle $4\pi$ SIT-like pulses in its center has been studied. In this case, the medium in the center remains constantly unexcited, while periodic inversion gratings and complex polarization standing waves are formed on either side in the region of the overlapping pulses. The spatial frequency of the inversion grating increases with the number of pulse collisions.

Outside the region of the pulse collisions, the medium is practically not excited, since in these regions the pulses propagate in the SIT regime. The result is a microcavity with Bragg-like mirrors. Its parameters can be controlled by the number of pulses. The results are undoubtedly interesting for the physics of ESP interaction with resonant media. Their coherent interaction with matter is poorly understood. The results of this research open up new avenues of investigation in ultrafast optics and attosecond physics and can be used for example, for the formation ultrafast optical switchers and other interesting applications [9].

**Funding.**
Russian Science Foundation, project 21-72-10028 (study of grating formation), project 23-12-00012 (analysis of gratings dependence on the parameters and their reflectivity calculation)

## References

1. H. J. Eichler, P. Günter, D. W. Pohl. Laser-induced dynamic gratings (Vol. 50). Springer (2013).
2. R. Kashyap. Fiber bragg gratings. Academic press (2009).
3. T. Jones, W. K. Peters, A. Efimov, D. Yarotski, R. Trebino, and P. Bowlan," Measuring an ultrashort, ultraviolet pulse in a slowly responding, absorbing medium," Opt. Express **29**, 11394 (2021).
4. R. M. Arkhipov," Electromagnetically induced gratings created by few-cycle light pulses (brief review)," JETP Lett. **113**, 611 (2021).
5. K. Midorikawa, "Progress on table-top isolated attosecond light sources," Nat. Photon. **16,** 267 (2022).
6. B. Xue, K. Midorikawa, and E. J. Takahashi, "Gigawatt-class, tabletop, isolated-attosecond-pulse light source," Optica **9**, 360 (2022).
7. M. T. Hassan, T. T. Luu, A. Moulet, O. Raskazovskaya, P. Zhokhov, M. Garg, N. Karpowicz, A. M. Zheltikov, V. Pervak, F. Krausz, E. Goulielmakis, "Optical attosecond pulses and tracking the nonlinear response of bound electrons," Nature **530**, 66 (2016).
8. D. Hui, H. Alqattan, S. Yamada, V. Pervak, K. Yabana, ,M. T. Hassan, "Attosecond electron motion control in dielectric," Nature Photonics **16**(1), 33 (2022).
9. M.T. Hassan, "Lightwave Electronics: Attosecond Optical Switching," ACS Photonics , https://doi.org/10.1021/acsphotonics.3c01584 (2024)
10. R. M. Arkhipov, M. V. Arkhipov, A. V. Pakhomov, P.A. Obraztsov, and N. N. Rosanov, "Unipolar and Subcycle Extremely Short Pulses: Recent Results and Prospects (Brief Review)," JETP Lett., 117(1) (2023).
11. H. Alqattan, D. Hui, V. Pervak, M.T. Hassan, "Attosecond light field synthesis," APL Photonics **7** (4) (2022).
12. H.-C. Wu and J. Meyer-ter Vehn, "Giant half-cycle attosecond pulses," Nature Photon. **6**, 304 (2012).
13. Q. Xin, Y. Wang, X. Yan, and B. Eliasson, "Giant isolated half-cycle attosecond pulses generated in coherent bremsstrahlung emission regime," Phys. Rev. E **107**, 035201 (2023).
14. R.M. Arkhipov, M.V. Arkhipov, I. Babushkin, A. Demircan, U. Morgner, N.N. Rosanov, " Ultrafast creation and control of population density gratings via ultraslow polarization waves,' Opt. Lett. **41**, 4983 (2016).
15. R.M. Arkhipov, M.V. Arkhipov, I. Babushkin, A. Demircan, U. Morgner. N.N. Rosanov, "Population density gratings induced by few-cycle optical pulses in a resonant medium," Scientific Reports **7**, 12467 (2017).
16. R. Arkhipov, A Pakhomov, M Arkhipov, I Babushkin, A Demircan, U. Morgner N.N. Rosanov, "Population difference gratings created on vibrational transitions by nonoverlapping subcycle THz pulses," Scientific Reports, **11** (1961) (2021).
17. R.M. Arkhipov, M.V. Arkhipov, N.N. Rosanov, "On the possibility of holographic recording in the absence of coherence between a reference beam and a beam scattered by an object," JETP Letters **111** (9), 484(2020).
18. R.M. Arkhipov, M.V. Arkhipov, A. V. Pakhomov, O.O. Diachkova, N.N. Rosanov, "Interference of the Electric and Envelope Areas of Ultrashort Light Pulses in Quantum Systems", Radiophysics and Quantum Electronics **66**(4), 286 (2023)
19. N.N. Rosanov, V.E. Semenov, N.V. Vyssotina, "Collisions of few-cycle dissipative solitons in active nonlinear fibers," Laser Physics 17, 1311 (2007).
20. M. Arkhipov, R. Arkhipov, I. Babushkin, N. Rosanov, "Self-stopping of light", Physical Review Letters **128** (20), 203901 (2022).
21. R.M. Arkhipov, M.V. Arkhipov, A.V. Pakhomov, O.O. Diachkova, N.N. Rosanov, "Nonharmonic Spatial Population Difference Structures Created by Unipolar Rectangular Pulses in a Resonant Medium," Opt. Spectr., **130** (11), 1443 (2022).
22. O.O. Diachkova, R.M. Arkhipov, M.V. Arkhipov, A.V. Pakhomov, N.N. Rosanov, "Light-induced dynamic microcavities created in a resonant medium by collision of non-harmonic rectangular 1-fs light pulses," Optics Communications, **538** , 129475 (2023).
23. R. M. Arkhipov, O. O. Diachkova, M. V. Arkhipov, A.V. Pakhomov., N.N. Rosanov, "Dynamics of microcavities created by nonharmonic unipolar light pulses in a resonant medium,"PREPRINT (Version 1) available at Research Square [https://doi.org/10.21203/rs.3.rs-3614406/v1], Applied Phys. B, accepted (2024)
24. S. L. McCall, E. L. Hahn, Phys. Rev. **183**, 457 (1969).
25. E. Galiffi, R. Tirole, S. Yin, H. Li, S. Vezzoli, P. A Huidobro, M. G. Silveirinha, R. Sapienza, A. Alù, J. B. Pendry, "Photonics of time-varying media," Adv. Phot. **4**(1) (2022) 014002.
26. Y. Sharabi, A. Dikopoltsev, E. Lustig, Y. Lumer, M.Segev, ,,Spatiotemporal photonic crystals,"Optica **9**(6) ,585 (2022).
27. A. Yariv. Quantum Electronics (Wiley, N.Y., 1975).
28. R. Arkhipov, M. Arkhipov, I. Babushkin, A. Pakhomov, N. Rosanov, "Coherent propagation of a half-cycle unipolar attosecond pulse in a resonant two-level medium," Journal of the Optical Society of America B **38** (6), 2004 (2021).
29. R. W. Boyd, Nonlinear Optics 3rd edn, (NY: Academic, 2008)
30. A. Othonos, K.Kalli, D.Pureur et al. *Fibre Bragg Gratings*. In: Wavelength Filters in Fibre Optics. Springer Series in Optical Sciences, vol 123. (Springer, Berlin, Heidelberg 2006).